\documentclass{PoS}

\title{Multiloop Euler-Heisenberg Lagrangians, Schwinger pair creation, and the QED N - photon amplitudes}

\ShortTitle{Multiloop EHL}

\author{Idrish Huet\\
      Facultad de Ciencias en F\'isica y Matem\'aticas, Universidad Aut\'onoma de Chiapas\\
 Ciudad Universitaria, Tuxtla Guti\'errez 29050, Mexico  \\
        E-mail: \email{idrish@ifm.umich.mx}}

       \author{Michel Rausch de Traubenberg\\
       IPHC-DRS, UdS, IN2P3,
23 rue du Loess, F-67037 Strasbourg Cedex, France \\
        E-mail: \email{Michel.Rausch@iphc.cnrs.fr}}
        
        \author{\speaker{Christian Schubert}\\
       Instituto de F{{\'\i}}sica y Matem\'aticas, Universidad Michoacana de San Nicol\'as de Hidalgo\\
Apdo. Postal 2-82, C.P. 58040, Morelia, Michoacan, Mexico \\
        E-mail: \email{schubert@ifm.umich.mx}}

\abstract{An update is given on our long-term effort to perform a three-loop check on the Affleck-Alvarez-Manton/Lebedev-Ritus
exponentiation conjecture for the imaginary part of the Euler-Heisenberg Lagrangian, using 1+1 dimensional QED as a toy model.
After reviewing the history and significance of the conjecture, 
we present trigonometric integral representations for the single electron loop contributions to the three-loop Lagrangian, and develop a symmetry-based
method for the calculation of their weak-field expansion coefficients.}

\FullConference{Loops and Legs in Quantum Field Theory (LL2018)\\
		29 April 2018 - 04 May 2018\\
		St. Goar, Germany}

\begin{document}

% MATH SYMBOLS
%
\def\cosech{\rm cosech}
\def\sech{\rm sech}
\def\coth{\rm coth}
\def\tanh{\rm tanh}
%fractions
\def\half{{1\over 2}}
\def\third{{1\over3}}
\def\fourth{{1\over4}}
\def\fifth{{1\over5}}
\def\sixth{{1\over6}}
\def\seventh{{1\over7}}
\def\eigth{{1\over8}}
\def\ninth{{1\over9}}
\def\tenth{{1\over10}}
\def\bN{\mathop{\bf N}}
\def\R{{\rm I\!R}}
\def\Eins{{\mathchoice {\rm 1\mskip-4mu l} {\rm 1\mskip-4mu l}
{\rm 1\mskip-4.5mu l} {\rm 1\mskip-5mu l}}}
\def\Z{{\mathchoice {\hbox{$\sf\textstyle Z\kern-0.4em Z$}}
{\hbox{$\sf\textstyle Z\kern-0.4em Z$}}
{\hbox{$\sf\scriptstyle Z\kern-0.3em Z$}}
{\hbox{$\sf\scriptscriptstyle Z\kern-0.2em Z$}}}}
\def\abs#1{\left| #1\right|}
\def\com#1#2{
        \left[#1, #2\right]}
\def\square{\kern1pt\vbox{\hrule height 1.2pt\hbox{\vrule width 1.2pt
   \hskip 3pt\vbox{\vskip 6pt}\hskip 3pt\vrule width 0.6pt}
   \hrule height 0.6pt}\kern1pt}
      \def\boxop{{\raise-.25ex\hbox{\square}}}
% \contract is a differential geometry contraction sign _|
\def\contract{\makebox[1.2em][c]{
        \mbox{\rule{.6em}{.01truein}\rule{.01truein}{.6em}}}}
\def\ltap{\ \raisebox{-.4ex}{\rlap{$\sim$}} \raisebox{.4ex}{$<$}\ }
\def\gtap{\ \raisebox{-.4ex}{\rlap{$\sim$}} \raisebox{.4ex}{$>$}\ }
\def\mn{{\mu\nu}}
\def\rs{{\rho\sigma}}
\newcommand{\Det}{{\rm Det}}
\def\Tr{{\rm Tr}\,}
\def\tr{{\rm tr}\,}
\def\sumij{\sum_{i<j}}
\def\e{\,{\rm e}}
%boldface vectors
\def\non{\nonumber\\}
\def\br{{\bf r}}
\def\bp{{\bf p}}
\def\bx{{\bf x}}
\def\by{{\bf y}}
\def\brhat{{\bf \hat r}}
\def\bv{{\bf v}}
\def\ba{{\bf a}}
\def\bE{{\bf E}}
\def\bB{{\bf B}}
\def\bA{{\bf A}}
%derivatives
\def\pa{\partial}
\def\dA{\partial^2}
\def\ddx{{d\over dx}}
\def\ddt{{d\over dt}}
\def\der#1#2{{d #1\over d#2}}
\def\lie{\hbox{\it \$}} % fancy L for the Lie derivative
\def\partder#1#2{{\partial #1\over\partial #2}}
\def\secder#1#2#3{{\partial^2 #1\over\partial #2 \partial #3}}
%
%equations
%\newcommand{\be}{\begin{equation}}
%\newcommand{\ee}{\end{equation}\noindent}
%\newcommand{\bear}{{\begin{eqnarray}}}
%\newcommand{\ear}{{\end{eqnarray}\noindent}}
%\newcommand{\benn}{\begin{enumerate}}
%\newcommand{\enn}{\end{enumerate}}
%\newcommand{\veject}{\vfill\eject}
%\newcommand{\ven}{\vfill\eject\noindent}
\def\be{\begin{equation}}
\def\ee{\end{equation}\noindent}
\def\bear{\begin{eqnarray}}
\def\ear{\end{eqnarray}\noindent}
\def\bec{\begin{equation}}
\def\eec{\end{equation}\noindent}
\def\bearc{\begin{eqnarray}}
\def\earc{\end{eqnarray}\noindent}
\def\benn{\begin{enumerate}}
\def\enn{\end{enumerate}}
\def\veject{\vfill\eject}
\def\ven{\vfill\eject\noindent}
%
%reference to equations
\def\eq#1{{eq. (\ref{#1})}}
\def\eqs#1#2{{eqs. (\ref{#1}) -- (\ref{#2})}}
%
%integrals
\def\totint{\int_{-\infty}^{\infty}}
\def\posint{\int_0^{\infty}}
\def\negint{\int_{-\infty}^0}
\def\pint{{\dps\int}{dp_i\over {(2\pi)}^d}}
%
% PHYS SYMBOLS
\newcommand{\GeV}{\mbox{GeV}}
\def\FFdual{F\cdot\tilde F}
\def\bra#1{\langle #1 |}
\def\ket#1{| #1 \rangle}
\def\braket#1#2{\langle {#1} \mid {#2} \rangle}
\def\vev#1{\langle #1 \rangle}
\def\rightvac{\mid 0\rangle}
\def\leftvac{\langle 0\mid}
\def\ihbar{{i\over\hbar}}
% dirac matrix stuff
\def\ge{\hbox{$\gamma_1$}}
\def\gz{\hbox{$\gamma_2$}}
\def\gd{\hbox{$\gamma_3$}}
\def\go{\hbox{$\gamma_1$}}
\def\gt{\hbox{\$\gamma_2$}}
\def\gth{\hbox{$\gamma_3$}} 
\def\gf{\hbox{$\gamma_5\;$}}
\def\slash#1{#1\!\!\!\raise.15ex\hbox {/}}
\newcommand{\slD}{\,\raise.15ex\hbox{$/$}\kern-.27em\hbox{$\!\!\!D$}}
\newcommand{\slpartial}{\raise.15ex\hbox{$/$}\kern-.57em\hbox{$\partial$}}
\newcommand{\PP}{\cal P}
\newcommand{\G}{{\cal G}}
\newcommand{\nc}{\newcommand}
\newcommand{\Fkala}{F_{\kappa\lambda}}
\newcommand{\Fkanu}{F_{\kappa\nu}}
\newcommand{\Flaka}{F_{\lambda\kappa}}
\newcommand{\Flamu}{F_{\lambda\mu}}
\newcommand{\Fmunu}{F_{\mu\nu}}
\newcommand{\Fnumu}{F_{\nu\mu}}
\newcommand{\Fnuka}{F_{\nu\kappa}}
\newcommand{\Fmuka}{F_{\mu\kappa}}
\newcommand{\Fkalamu}{F_{\kappa\lambda\mu}}
\newcommand{\Flamunu}{F_{\lambda\mu\nu}}
\newcommand{\Flanumu}{F_{\lambda\nu\mu}}
\newcommand{\Fkamula}{F_{\kappa\mu\lambda}}
\newcommand{\Fkanumu}{F_{\kappa\nu\mu}}
\newcommand{\Fmulaka}{F_{\mu\lambda\kappa}}
\newcommand{\Fmulanu}{F_{\mu\lambda\nu}}
\newcommand{\Fmunuka}{F_{\mu\nu\kappa}}
\newcommand{\Fkalamunu}{F_{\kappa\lambda\mu\nu}}
\newcommand{\Flakanumu}{F_{\lambda\kappa\nu\mu}}

\newcommand{\pb}{\bar{p}}
\newcommand{\ph}{\hat{p}}
\newcommand{\gb}{\bar{g}}
\newcommand{\gh}{\hat{g}}
\newcommand{\zb}{\bar{z}}
\newcommand{\zh}{\hat{z}}
\newcommand{\wh}{\hat w}
\newcommand{\wb}{\bar w}
\newcommand{\p}{p\!\!\!/~}
\newcommand{\pbdash}{\bar{p} \!\!\!/~}
\newcommand{\q}{q\!\!\!/~}
\newcommand{\B}{\beta \!\!\!/~}
\newcommand{\tb}{\bar{t}}

\nc{\spa}[3]{\left\langle#1\,#3\right\rangle}
\nc{\spb}[3]{\left[#1\,#3\right]}
\nc{\ksl}{\not{\hbox{\kern-2.3pt $k$}}}
\nc{\hf}{\textstyle{1\over2}}
\nc{\pol}{\varepsilon}
\nc{\tq}{{\tilde q}}
\nc{\esl}{\not{\hbox{\kern-2.3pt $\pol$}}}
\newcommand{\cL}{\cal L}
\newcommand{\D}{\cal D}
\newcommand{\Dhalf}{{D\over 2}}
\def\eps{\epsilon}
\def\epshalf{{\epsilon\over 2}}
\def\lag{( -\partial^2 + V)}
%worldline
\def\freeexp{{\rm e}^{-\int_0^Td\tau {1\over 4}\dot x^2}}
\def\kinb{{1\over 4}\dot x^2}
\def\kinf{{1\over 2}\psi\dot\psi}
\def\expk{{\rm exp}\biggl[\,\sum_{i<j=1}^4 G_{Bij}k_i\cdot k_j\biggr]}
\def\expp{{\rm exp}\biggl[\,\sum_{i<j=1}^4 G_{Bij}p_i\cdot p_j\biggr]}
\def\expshort{{\e}^{\half G_{Bij}k_i\cdot k_j}}
\def\expabb{{\e}^{(\cdot )}}
\def\epseps#1#2{\varepsilon_{#1}\cdot \varepsilon_{#2}}
\def\epsk#1#2{\varepsilon_{#1}\cdot k_{#2}}
\def\kk#1#2{k_{#1}\cdot k_{#2}}
\def\G#1#2{G_{B#1#2}}
\def\Gp#1#2{{\dot G_{B#1#2}}}
\def\GF#1#2{G_{F#1#2}}
\def\Dab{{(x_a-x_b)}}
\def\Dsq{{({(x_a-x_b)}^2)}}
\def\PITD{{(4\pi T)}^{-{D\over 2}}}
\def\4piTD{{(4\pi T)}^{-{D\over 2}}}
\def\4piT4{{(4\pi T)}^{-2}}
\def\TintmD{{\dps\int_{0}^{\infty}}{dT\over T}\,e^{-m^2T}
    {(4\pi T)}^{-{D\over 2}}}
\def\Tintm4{{\dps\int_{0}^{\infty}}{dT\over T}\,e^{-m^2T}
    {(4\pi T)}^{-2}}
\def\Tintm{{\dps\int_{0}^{\infty}}{dT\over T}\,e^{-m^2T}}
\def\Tint{{\dps\int_{0}^{\infty}}{dT\over T}}
\def\np{n_{+}}
\def\nm{n_{-}}
\def\Np{N_{+}}
\def\Nm{N_{-}}
\newcommand{\slG}{{{\dot G}\!\!\!\! \raise.15ex\hbox {/}}}
\newcommand{\Gd}{{\dot G}}
\newcommand{\Gund}{{\underline{\dot G}}}
\newcommand{\Gdd}{{\ddot G}}
\def\GBd12{{\dot G}_{B12}}
\def\Dx{\dps\int{\cal D}x}
\def\Dy{\dps\int{\cal D}y}
\def\Dpsi{\dps\int{\cal D}\psi}
\def\dint#1{\int\!\!\!\!\!\int\limits_{\!\!#1}}
\def\ddtau{{d\over d\tau}}
\def\ie{\hbox{$\textstyle{\int_1}$}}
\def\iz{\hbox{$\textstyle{\int_2}$}}
\def\id{\hbox{$\textstyle{\int_3}$}}
\def\ldop{\hbox{$\lbrace\mskip -4.5mu\mid$}}
\def\rdop{\hbox{$\mid\mskip -4.3mu\rbrace$}}
%
%VARIOUS
\newcommand{\1}{{\'\i}}
\newcommand{\no}{\noindent}
\def\non{\nonumber}
\def\dps{\displaystyle}
\def\sy{\scriptscriptstyle}
\def\sy{\scriptscriptstyle}

\section{Euler-Heisenberg Lagrangian and photon amplitudes}

In 1936 Heisenberg and Euler \cite{eulhei} calculated what nowadays is called the one-loop QED
effective Lagrangian in a constant field (`` Euler-Heisenberg Lagrangian'' = EHL), obtaining the following well-known
integral representation:

\bear
{\cal L}^{(1)}(a,b)&=& - {1\over 8\pi^2}
\int_0^{\infty}{dT\over T^3}
\,\e^{-m^2T}
\biggl\lbrack
{(eaT)(ebT)\over {\rm tanh}(eaT)\tan(ebT)} 
%\nonumber\\&&\hspace{70pt}
- {e^2\over 3}(a^2-b^2)T^2 -1
\biggr\rbrack
\, .
\nonumber
\label{ehspin}
\ear
Here  $a,b$  are the two
invariants of the Maxwell field, 
related to $\bf E$, $\bf B$ by $a^2-b^2 = B^2-E^2,\quad ab = {\bf E}\cdot {\bf B}$.
The superscript $(1)$ stands for one-loop. 
A similar representation was obtained shortly later for scalar QED by Weisskopf \cite{weisskopf}.

\begin{figure}[htbp]
\begin{center}
\includegraphics[scale=.6]{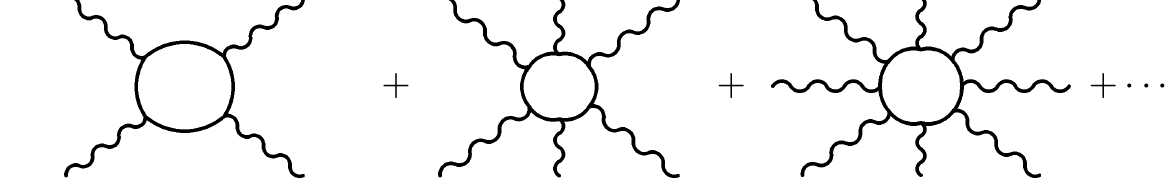}
%\caption{{\bf }}
\end{center}
\caption{Sum of diagrams equivalent to the one-loop EHL}
\label{fig-EHL1loop}
\end{figure}

The EHL holds the information on
the $N$ - photon amplitudes in the low energy limit, where all photon energies are small
compared to the electron mass, $\omega_i\ll m$ .  
Diagrammatically, this corresponds to Fig. \ref{fig-EHL1loop}

In \cite{56} it was shown how to construct these amplitudes explicitly 
from the weak field expansion coefficients  $c_{kl}$,  defined by

\bear
{\cal L}^{(1)} (a,b) = \sum_{k,l} c_{kl}\, a^{2k}b^{2l} \, .
\ear
In particularly, there it was shown that,  
for each $ N$ and each given helicity assignment, the dependence on the momentum and polarization vectors can be absorbed into a  single invariant $ \chi_N$.

\section{Imaginary part and Sauter-Schwinger pair creation}

\begin{figure}[h]
%\begin{picture}(0,0)(6000,10000)
\centerline{\centering
\includegraphics[scale=.7]{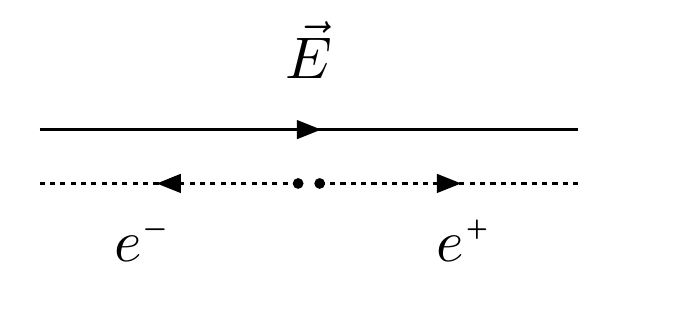}
}
%\end{picture}
\caption{Pair creation by an external field.}
\label{fig-pairtunnel}
\end{figure}

If the field has an electric component ($ b \ne 0$) then there are poles on the 
integration contour at $ ebT =  k\pi$ which create an imaginary part.
For the purely electric case one gets \cite{schwinger51}

\begin{eqnarray}
{\rm Im} {\cal L}^{(1)}(E) &=&  \frac{m^4}{8\pi^3}
\beta^2\, \sum_{k=1}^\infty \frac{1}{k^2}
\,\exp\left[-\frac{\pi k}{\beta}\right]
%\non\\
%{\rm Im}{\cal L}_{\rm scal}^{(1)}(E) 
%&=&
%-\frac{m^4}{16\pi^3}
%\beta^2\, \sum_{k=1}^\infty \frac{(-1)^{k}}{k^2}
%\,\exp\left[-\frac{\pi k}{\beta}\right]
\label{schwinger}
\end{eqnarray}
($\beta = eE/m^2$). Physically, in this decomposition 
the $ k$th term relates to  coherent creation of $ k$  pairs in
one Compton volume. 
In the following we will consider only the weak-field limit  $\beta \ll 1$, where only the leading $k=1$ is relevant. 
Note that $ {\rm Im}{\cal L}^{(1)}(E)$ depends on $ E$  non-perturbatively (nonanalytically), which is consistent with
the interpretation of pair creation as  vacuum tunneling, originally due to Sauter, where a virtual electron-positron pair turns real by 
extracting their rest mass energies from the external field (Fig. \ref{fig-pairtunnel}). 

\begin{figure}[ht]
\centerline{\hspace{30pt}\includegraphics[scale=.5]{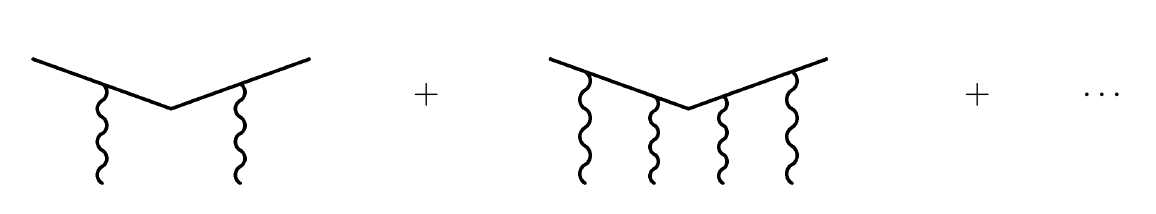}}
%   \includegraphics{fig1.pdf}
%\end{figure}

\no
\caption{``Cut'' diagrams describing Schwinger pair creation.}
\label{fig-schwinger_tree_diags}
\end{figure}

\no
This connection between the effective action and the pair creation rate is based on the  Optical Theorem, which relates
the imaginary part of the diagrams shown in Fig. \ref{fig-EHL1loop} to the `` cut diagrams '' shown in Fig. \ref{fig-schwinger_tree_diags}.

\no
However, the latter individually all vanish for a constant field, which can emit only zero-energy photons.
Thus for a constant field we cannot use dispersion relations for individual diagrams;
what counts is rather the asymptotic behaviour of the diagrams for a large number of photons. 
The appropriate generalization is then a Borel dispersion relation. This works in the following way \cite{37}:
define the  weak field expansion coefficients of the EHL by

\bear
{\cal L}^{(1)}(E) = \sum_{n=2}^{\infty} c(n) \Bigl(\frac{eE}{m^2}\Bigr)^{2n} \, .
\ear
It can be shown that their leading large - $n$ behavior is

\bear
c(n)\,\, {\stackrel{n\to \infty}{\sim}} \,\,c_{\infty} \Gamma[2n - 2] \, .
\label{leading}
\ear
The Borel dispersion relation relates this leading behavior to the leading weak-field behavior
of the imaginary part of the Lagrangian:

\bear
{\rm Im}{\cal L}^{(1)}(E) \, \, {\stackrel{\beta\to 0}{\sim}} \,\,
c_{\infty}\,\e^{-\frac{\pi m^2}{eE}} \, .
\ear
Thus we have rederived the leading Schwinger exponential of (\ref{schwinger}) in a way that might seem rather indirect and
complicated. However, this approach turns out to be very useful for higher-loop considerations.

\section{Beyond one loop}

\begin{figure}[h]
%\begin{picture}(0,0)(6000,10000)
\centerline{\centering
\includegraphics[scale=.6]{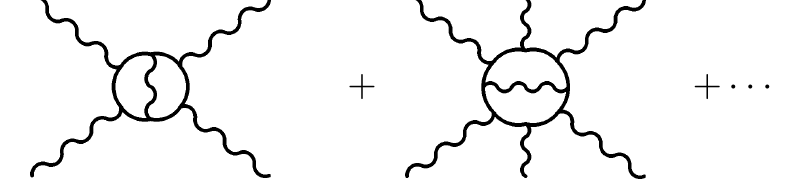}
}
\caption{Feynman diagrams correposnding to the 2-loop EHL.}
\label{fig-2loopEHL}
\end{figure}

The two-loop (one-photon exchange) correction to the EHL corresponds to the set of diagrams shown in Fig. 
\ref{fig-2loopEHL} (there is also a one-particle reducible contribution \cite{giekar}, but for our present purposes it can be discarded).

\begin{figure}[h]
\centerline{\centering
\includegraphics[scale=.3]{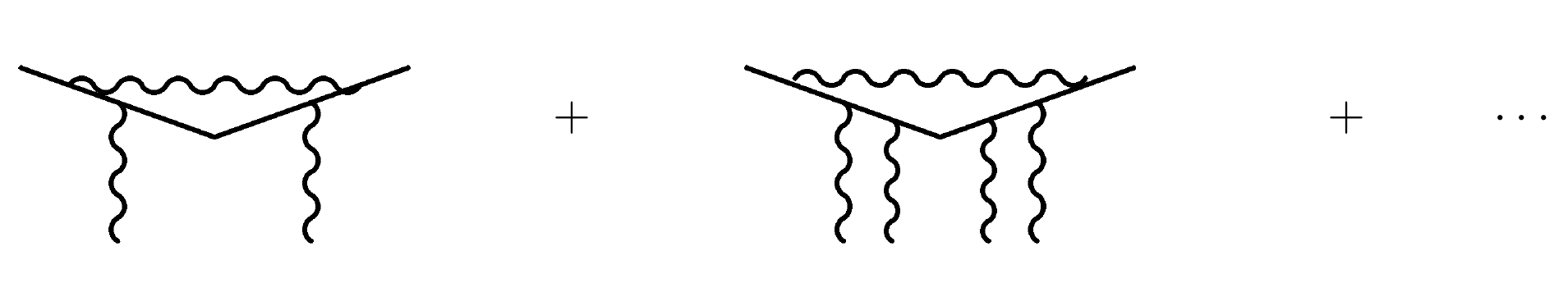}
}
\caption{Feynman diagrams contributing to 2-loop Schwinger pair creation.}
\label{fig-schwinger_tree_adorned_diags}
\end{figure}

The corresponding corrections to the tree-level pair creation diagrams of Fig. \ref{fig-schwinger_tree_diags} are shown in Fig. 
\ref{fig-schwinger_tree_adorned_diags}.

Even at the two-loop level, the study of the EHL has already a quite substantial history 
\cite{ritusspin,lebrit,ditreu-book,18}. 
Unfortunately, it leads to a type of rather intractable two-parameter integrals.
However, the imaginary part $ {\rm Im}{\cal L}^{(2)}(E)$ admits a decomposition similar to Schwinger's one-loop one, eq. (\ref{schwinger}),
and in the weak-field limit it becomes a simple addition to it \cite{lebrit}:
 
 \bear
{\rm Im} {\cal L}^{(1)} (E) +
{\rm Im}{\cal L}^{(2)} (E) 
\,\,\,\, {\stackrel{\beta\to 0}{\sim}} \,\,\,\,
 \frac{m^4\beta^2}{8\pi^3}
\bigl(1+\alpha\pi\bigr)
\,{\rm e}^{-{\pi\over\beta}}
\, .
\non
\label{Im1plus2}
\ear
\medskip

\no
In \cite{lebrit}, Lebedev and Ritus further noted that, if one assumed that higher orders will lead to exponentiation,

 \bear
{\rm Im} {\cal L}^{(1)} (E) +
{\rm Im}{\cal L}^{(2)} (E) 
+
{\rm Im}{\cal L}^{(3)} (E) + \ldots
\,\,\,\, {\stackrel{\beta\to 0}{\sim}} \,\,\,\,
 \frac{m^4\beta^2}{8\pi^3}
\,{\rm exp}\Bigl[ -{\pi\over\beta}+\alpha\pi \Bigr]
= {\rm Im}{\cal L}^{(1)}(E)\,\,{\rm e}^{\alpha\pi}
 \nonumber
\ear
then the result could be interpreted in the tunneling picture as the  corrections to the
Schwinger pair creation rate due to the pair being created with a negative Coulomb
interaction energy. This lowers the energy that has to be drawn from the field, and
can be interpreted as a mass shift

\bear
m(E) \approx m + \delta m(E), \quad \delta m(E) =  - \frac{\alpha}{2}\frac{eE}{m} 
\nonumber
\ear
where $\delta m (E)$ is just the ``Ritus mass shift'', originally derived from the
crossed process of one-loop electron propagation in the field \cite{ritusmass}. 

% (Fig. \ref{fig-crossing}). 
%
%\vspace{-5pt}
%
%\begin{figure}[h]
%%\begin{minipage}[b]{0.5\linewidth}
%\centering
%\hspace{10pt}\includegraphics[scale=.5]{fig-uncrossed.pdf}
%\raisebox{3.1 em} {$\hspace{20pt}\Longleftrightarrow\hspace{20pt}$}
%\includegraphics[scale=.4]{fig-crossed.pdf}
%\hspace{20pt}
%\label{fig4point}
%%\end{minipage}
%%\hspace{0cm}
%%\begin{minipage}[b]{0.5\linewidth}
%%\centering
%%\includegraphics[scale=.8]{fighexagon.pdf}
%\caption{Crossing relation between pair creation and electron propagation in the field.}
%\label{fig-crossing}
%%\end{minipage}
%\end{figure}
%%\hspace{70pt}  Pair creation by the field  
%%\hspace{30pt}  Electron propagation in the field

Unbeknownst to those authors, for  scalar QED the corresponding exponentiation had been conjectured already
two years earlier by Affleck, Alvarez and Manton \cite{afalma}:

\bear
\sum_{l=1}^{\infty}{\rm Im}{\cal L}^{(l)}_{\rm scal}(E)
&{\stackrel{\beta\to 0}{\sim}}&
 -\frac{m^4\beta^2}{16\pi^3}
\,{\rm exp}\Bigl[ -{\pi\over\beta}+\alpha\pi \Bigr] =  {\rm Im}{\cal L}^{(1)}_{\rm scal}(E)\,\,{\rm e}^{\alpha\pi} \, .
\label{ImLallloop}
\nonumber
\ear
However, they arrived at this conjecture in a very different way, namely
using Feynman's  worldline path integral formalism in a semi-classical
approximation ( ``worldline instanton''). 

\begin{figure}[h]
%\begin{picture}(0,0)(6000,10000)
\centerline{\centering
\includegraphics[scale=.5]{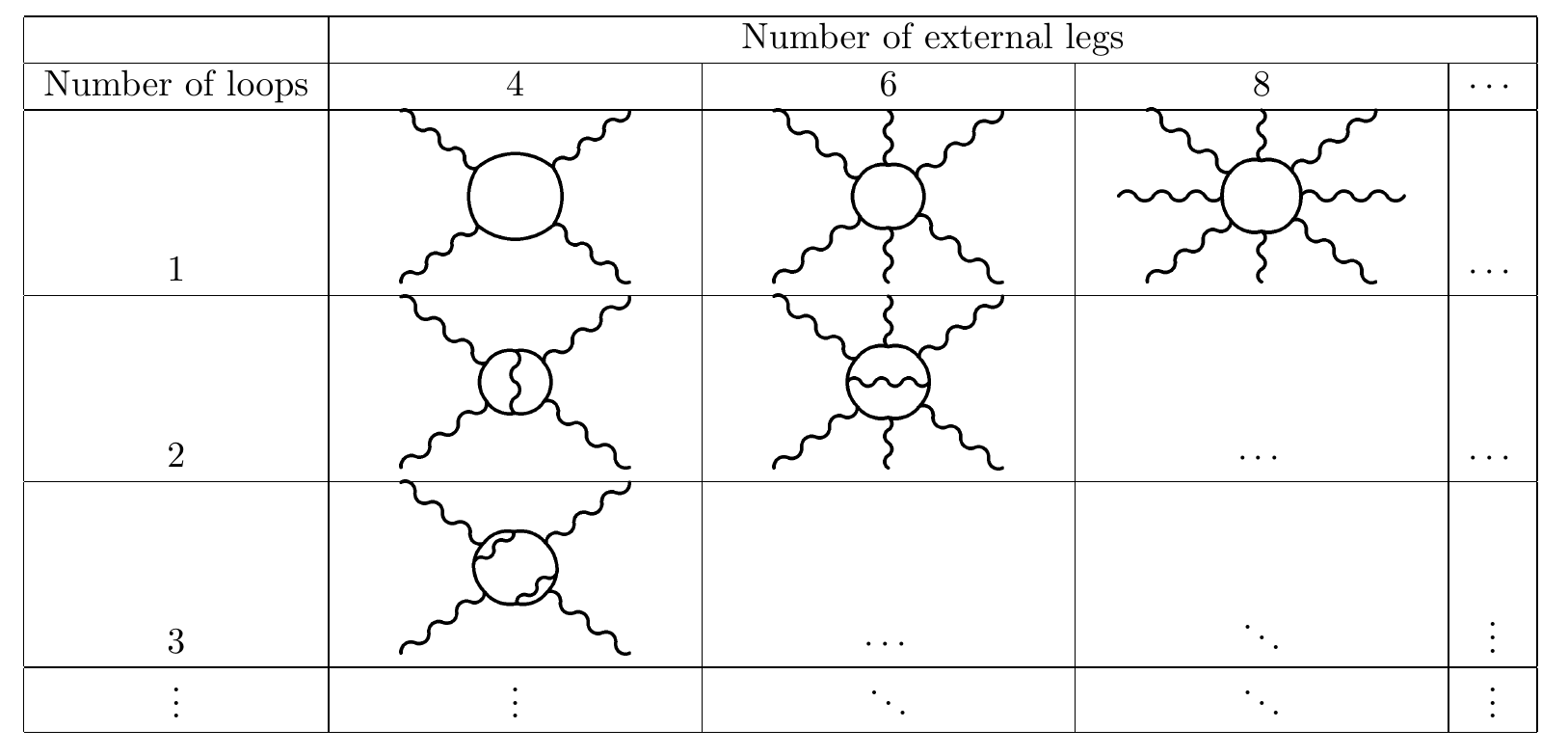}
}
\caption{Feynman diagrams contributing to the AAM formula.}
\label{fig-aamdiag}
\end{figure}

Thus, although neither derivation is rigorous, there is much that speaks for the exponentiation conjecture. 
If true, it would constitute a unique example of a double summation of an infinite set of Feynman diagrams, 
involving any number of loops and legs, depicted in Fig. \ref{fig-aamdiag}.

Here it is understood that the ``horizontal'' summation is performed using the leading large $n$ approximation for the
weak-field expansion coefficients, which at each fixed loop order produces the same leading Schwinger exponential $ {\rm e}^{-\frac{\pi}{\beta}}$.
The ``vertical'' summation over an increasing number of internal photon insertions produces the 
Affleck-Alvarez-Manton/Lebedev-Ritus factor $ {\rm e}^{\alpha\pi}$. Note that diagrams with more than one electron loop
are nor included, since they get suppressed in the weak-field/large $n$ limit. On the other hand, the counterdiagrams
from mass renormalization, although not shown here, have to be included. 
What is very surprising about the ``vertical'' exponentiation is, that it has produced the analytic factor $ {\rm e}^{\alpha\pi}$!
This is counter-intuitive, since the growth in the number of diagrams caused by the insertion of an increasing number of photons into an electron loop 
would lead one to expect a vanishing radius of convergence in $\alpha$. 

Let us mention also that, using Borel analysis, this factor can be transferred from the imaginary part of the effective Lagrangian to the large - $N$ limit of the $N$ - photon amplitudes
with all ``+'' polarizations \cite{60}: 

\bear
{\rm lim}_{N\to\infty}{\Gamma^{({\rm all-loop})}
[k_1,\varepsilon_1^+;\ldots ;k_N,\varepsilon_N^+]
\over 
\Gamma^{(1)}
[k_1,\varepsilon_1^+;\ldots ;k_N,\varepsilon_N^+]}
=
\e^{\alpha\pi}
\, .
\nonumber
\label{conj}
\ear
Here an essential ingredient is the above-mentioned fact that, independently of the loop order, the complete dependence of the $N$ - photon amplitudes
for a fixed helicity assignment can be absorbed into a single invariant.

\section{The EHL in 1+1 dimensional QED}

The exponentiation conjecture has been verified at the two-loop order by explicit computation in both scalar and spinor QED. 
A three-loop check is in order, but calculating the three-loop EHL in $ D=4$ seems presently technically out of reach. 
In 2005, Krasnansky \cite{krasnansky} studied the EHL for scalar QED in various spacetime dimensions. 
In 1+1 dimensions, he found the following explicit result for this Lagrangian:

\bear
{\cal L}_{\rm scal}^{(2)(2D)}(\kappa)
&=&
-\frac{e^2}{32\pi^2}\left(
\xi^2_{2D} 
-4\kappa \xi_{2D}'\right) ,
\nonumber
\ear
where $\xi_{2D}= -\Bigl(\psi(\kappa+\half)-\ln (\kappa)\Bigr)$,
$\psi(x)=\Gamma^\prime(x)/\Gamma(x)$, $\kappa =  m^2/(2ef)$, 
$f^2=\fourth F_{\mu\nu}F^{\mu\nu}$.

This is simpler than in four dimensions, but still non-trivial (in fact very similar in structure to the EHL
in four dimensions for a self-dual field \cite{51}), which suggests to use 1+1 dimensional QED as a toy model for 
testing the exponentiation conjecture.  
In \cite{81} two of the authors and G. McKeon used the method of \cite{afalma} to generalize the exponentiation conjecture to the 2D case,
in the following form:

\bear
{\rm Im}{\cal L}^{(all-loop)}_{2D}
\sim
\e^{-\frac{m^2\pi}{eE} + \tilde\alpha \pi^2  \kappa^2}
\label{2Dexp}
\ear
where $\tilde\alpha = \frac{2e^2}{\pi m^2}$ is our definition of the fine-structure constant in 2D.
There we also calculated the one- and two-loop contributions to the $2D$ EHL, obtaining
(dropping now the subscript `2D')

\bear
{\cal L}^{(1)}(\kappa ) &=& -{m^2\over 4\pi} {1\over\kappa}
\Bigl[{\rm ln}\Gamma(\kappa) - \kappa(\ln \kappa -1) +
\half \ln \bigl({\kappa\over 2\pi}\bigr)\Bigr] \, ,
\nonumber\\
{\cal L}^{(2)}(\kappa) &=& {m^2\over 4\pi}\frac{\tilde\alpha}{4}
\Bigl[ \tilde\psi(\kappa) + \kappa \tilde\psi'(\kappa)
+\ln(\lambda_0 m^2) + \gamma + 2 \Bigr] \, ,
\nonumber
\ear \no
where $\tilde\psi (x) \equiv \psi(x) - \ln x + {1\over 2x}$.
This allowed us to obtain explicit formulas not only for $ c^{(1)}(n)$ but also for $ c^{(2)}(n)$:

\bear
c^{(1)}(n) &=& (-1)^{n+1} \frac{B_{2n}}{4n(2n-1)}\, ,\nonumber\\
c^{(2)}(n) &=& (-1)^{n+1} \frac{\tilde\alpha}{8}\frac{2n-1}{2n}B_{2n} \, .\nonumber
\ear
Using properties of the Bernoulli numbers $ B_n$, it was then easy to verify the following prediction made
by the exponentiation conjecture for the ratio between the two-loop and the one-loop expansion coefficients:

\bear
\lim_{n\to\infty}  {c^{(2)}(n)\over c^{(1)}(n+1)} 
&=&
\tilde\alpha \pi^2
\nonumber
\ear
(the relative shift in the argument of the coefficients is due to the fact that, unlike the four-dimensional case, 
in two dimensions the term involving $\tilde\alpha$ in the exponent in (\ref{2Dexp}) also involves the external field).
The convergence of $ c^{(2)}(n)$ to the asymptotic prediction is rather fast (Fig. \ref{fig-ratio2c1}):

\begin{figure}[h]
%\begin{picture}(0,0)(6000,10000)
\centerline{\centering
\includegraphics[scale=.55]{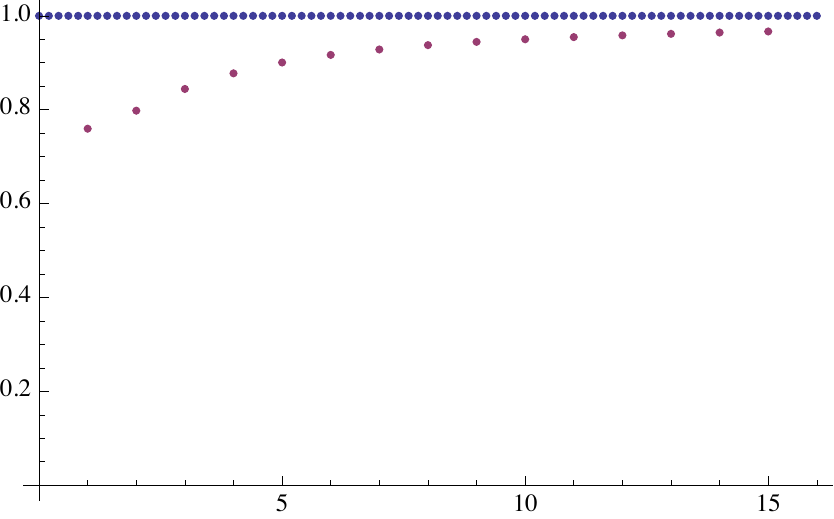}
}
\caption{Convergence of the two-loop coefficients to the asymptotic prediction.}
\label{fig-ratio2c1}
\end{figure}
\bigskip

\section{The three-loop EHL in 1 + 1 dimensional QED}

\begin{figure}[h]
%\begin{picture}(0,0)(6000,10000)
\centerline{\centering
\includegraphics[scale=.9]{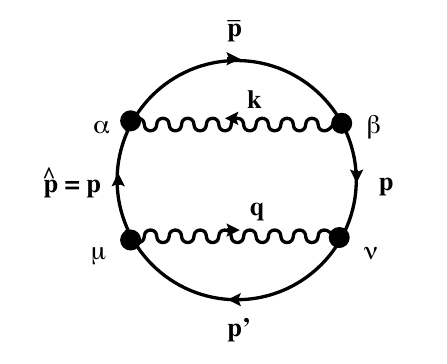}
\raisebox{-.6 em}
{\includegraphics[scale=.9]{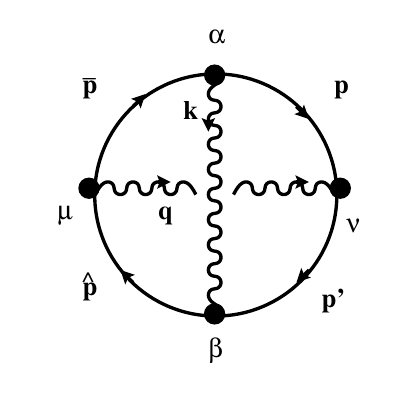}}
}
\hspace{155pt} A \hspace{101pt} B
\caption{Single electron-loop contributions to the three-loop EHL.}
\label{fig-AB}
\end{figure}

At the three-loop level, the calculation of the EHL becomes
challenging even in the two-dimensional
case. There are five different topologies of diagrams, but 
for the exponentiation conjecture only the diagrams with a single electron loop are relevant, 
depicted in Fig. \ref{fig-AB}.

We note that, due to the super-renormalizability of two-dimensional QED,  at three loops the 
EHL is already UV finite.
There are spurious IR - divergences, but we found that they can be removed by going to 
the ``traceless'' or ``anti-Feynman'' gauge $\xi = -1$. Although this gauge is known to have
some special properties (see \cite{gakamo} and A. Kataev's talk at this workshop) it is not clear
to us why this is the case. 
Both diagrams lead to four-parameter integrals with a trigonometric integrand. The one for diagram A is simple:

\begin{eqnarray}
{\cal L}^{3A} (f)
&=& \frac{\tilde\alpha^2m^2}{32\pi} \int_{0}^{\infty}dw dw'  d\hat w d\bar w~ I_A ~ \e^{-a} \, ,\nonumber\\
I_A &=& \frac{\rho^3}{A^2 \cosh \rho w \cosh \rho \hat w (\cosh \rho \bar w \cosh \rho w')^2} \Bigg[ \frac{\cosh \rho (w-\hat w)}{2\rho}  - \frac{1}{A \cosh \rho w  \cosh \rho \hat w} \Bigg]
\, ,
\nonumber
\end{eqnarray}
where 
$ \rho = \frac{ef}{m^2}, a = w+ w' + \hat w + \bar w, A = \tanh \rho w + \tanh \rho w' + \tanh \rho \hat w + \tanh \rho \bar w$.
The calculation of its contribution to the weak-field expansion coefficients $ c^{(3)A}(n) = \frac{\tilde\alpha^2}{64}\Gamma^A_n$
is straightforward, and yields rational numbers. The first few are

\bear
\Gamma^A_0 = -\frac{1}{3}, \Gamma^A_1 =  -\frac{1}{30}, \Gamma^A_2 = \frac{17}{63}, \Gamma^A_3 = \frac{251}{99}, \ldots
\ear
(so far we have obtained 13 coefficients).  As one would expect, the nonplanar diagram B leads to a more complicated integrand:

\begin{eqnarray}
{\cal L}^{3B} (f)
&=& \frac{\tilde\alpha^2m^2}{128\pi} \int_{0}^{\infty}dw dw' d\hat w d\bar w ~ I_B ~ \e^{-a} \, ,\nonumber\\
I_B &=& 
\frac{\rho^3}{\cosh^2 \rho w \cosh^2 \rho w' \cosh^2 \rho \hat w\cosh^2 \rho \bar w}
\frac{B}{A^3C} 
\nonumber\\&&
- \rho \frac{\cosh(\rho \tilde{w}) }{\cosh \rho w \cosh \rho w' \cosh \rho \hat w\cosh \rho \bar w} 
\Bigl\lbrack 
\frac{1}{A} - \frac{C}{G^2}\ln \Bigl(1+ \frac{G^2}{AC}\Bigr)\Bigr\rbrack \, ,
\nonumber
\label{Bw}
\end{eqnarray}

\bear
%A &=& \tanh z + \tanh z' + \tanh \hat z + \tanh \bar z \non
B &=& (\tanh^2 z + \tanh^2 \hat z)(\tanh z' + \tanh \bar z) +  (\tanh^2 z' + \tanh^2 \bar z)(\tanh z + \tanh \hat z) \, , \non\\
C &=& \tanh z\,\tanh z' \,\tanh\, \hat z + \tanh z\, \tanh z'\, \tanh \bar z + \tanh z\, \tanh \hat z\, \tanh \bar z + \tanh z'\, \tanh \hat z \,\tanh \bar z \, ,\non\\
%E &=& (\tanh z + \tanh z') (\tanh z + \tanh \bar z)(\tanh \hat z + \tanh z' ) (\tanh \hat z + \tanh \bar z) \non
G &=& \tanh z\, \tanh \hat z - \tanh z'\, \tanh \bar z \nonumber
\ear
($ z = \rho w$  etc.).
For diagram $ B$, the calculation of the weak-field expansion coefficients turned out to be much more difficult than for $ A$.
This is not only because the integrals are of a more difficult structure, and this time produce also $\zeta_3$ values, but 
also because the expansion in the external field creates huge numerator polynomials in the Feynman parameters. 
In a first attempt using numerical integration \cite{111} we obtained only six coefficients, which is much too few for our purposes. 
In a forthcoming paper, we show how to use the high symmetry of diagram B to solve both problems. 
For obtaining a first integral, we introduce the operator

\bear
\tilde d & \equiv & \frac{\partial}{\partial w}- \frac{\partial}{\partial w'}+ \frac{\partial}{\partial \hat w} -\frac{\partial}{\partial \bar w} 
\nonumber
\ear
which acts simply on the building blocks of the integrand. 
Integrating by parts with this operator, it is possible to write the integrand of the $n$-th coefficient $ \beta_n$ as a total derivative
$ \beta_n = \tilde d \theta_n$. Then, using again the symmetries of diagram B, 

\bear
\int_{0}^{\infty}dw dw' d\hat w  d\bar w \,\e^{-a} \beta_n &=& \int_{0}^{\infty}dw d\bar w d\hat w dw'
 \tilde d
   \,\e^{-(w+w'+\hat w +\bar w)}
\theta_n 
=
4\int_{0}^{\infty}dw dw' d\hat w \,\e^{-(w+w'+\hat w)}\, \theta_n\vert_{\bar w =0}
\, .
\nonumber
\ear
The remaining integrals are already of a standard type. In this way we obtained the first two coefficients:

\bear
\Gamma^B_0 &=& -\frac{3}{2}+\frac{7}{4}\zeta_3, \quad  \Gamma^B_1 = -\frac{251}{120} + \frac{35}{16}\zeta_3 \, .
\ear
We can also predict that all coefficients will be of the form $ r_1 + r_2 \zeta_3$ with rational numbers $ r_1,r_2$. 

For the purpose of obtaining more compact Feynman numerator polynomials, we note that Diagram $ B$ has the symmetries 

\bear
\qquad  w &\leftrightarrow& \hat w \nonumber\\
\quad w'  &\leftrightarrow& \bar w\nonumber\\
\quad (w,\hat w) &\leftrightarrow& (w',\bar w) \nonumber
\ear
Those generate the dihedral group $ D_4$. Using the theory of polynomial invariants of that group, this allows one to  rewrite the numerator polynomials 
as polynomials in the variable $ \tilde w \equiv w - w' +\hat w -\bar w$ with coefficients that are polynomials
in the four $ D_4$ - invariants $ a,v,j,h$,

\bear
a &=& w + w' + \hat w + \bar w \, , \nonumber\\
v &=& 2 (w \hat w + w' \bar w) + (w + \hat w)(w' + \bar w) \, , \nonumber\\
j &=& a \tilde w - 4 ( w \hat w - w' \bar w) \, ,\nonumber\\
h &=& a  (ww'\hat w + ww'\bar w + w\hat w \bar w + w'\hat w\bar w) + ( w \hat w - w' \bar w)^2 \, . \nonumber
\ear
These invariants are moreover chosen such that they are  annihiliated by $ \tilde d$.
Thus they are well-adapted to the integration-by-parts algorithm.
This  very significantly reduces the size of the expressions  generated by the expansion in the field.

\section{Outlook}

To summarize, we are confident to get enough expansion coefficients to settle the question of the validity of the exponentiation conjecture
in $1+1$ dimensional QED by the time of Loops and Legs 2020.
The techniques that we have developed for the calculation of the 3-loop EHL in 2D should also become useful in an eventual
calulation of this Lagrangian in four dimensions.

\end{document}